\documentclass[a4paper, 11pt]{article}

\usepackage{jheppub} %
\usepackage{graphicx}%
\usepackage{dcolumn}%
\usepackage{bm}%
\usepackage[utf8]{inputenc} %
\usepackage[T1]{fontenc}    %
\usepackage{url}            %
\usepackage{booktabs}       %
\usepackage{amsfonts}       %
\usepackage{microtype}      %
\usepackage{amssymb,amsmath,amsthm}
\usepackage{xcolor}
\usepackage{comment}
\usepackage{multirow}
\usepackage[caption=false]{subfig}
\usepackage{tabularx}
\usepackage{float}
\usepackage{algorithm}
\usepackage{algpseudocode}
\usepackage{enumitem} 
\usepackage{lipsum}
\usepackage{physics}
\usepackage{bbm}
\usepackage[pagewise]{lineno}
\usepackage{dsfont}

\begin{document}

\title{
 Simulating the dynamics of an SU(2) matrix model on a trapped-ion quantum computer
}%

\author[1]{Gavin S. Hartnett,}
\affiliation[1]{Q-CTRL, Los Angeles, CA USA and Sydney, NSW Australia}
\emailAdd{gavin.hartnett@q-ctrl.com}
\author[1]{Haoran Liao,}
\emailAdd{haoran.liao@q-ctrl.com}
\author[2,3,4]{Enrico Rinaldi}
\affiliation[2]{Quantinuum Ltd., Partnership House, Carlisle Place, London SW1P 1BX, UK}
\affiliation[3]{Center for Quantum Computing, RIKEN, 2-1 Hirosawa, Wako, Saitama 351-0198, Japan}
\affiliation[4]{School of Mathematical Sciences, Queen Mary University of London, Mile End Road, London, E1 4NS, UK}
\emailAdd{enrico.rinaldi@quantinuum.com}

\abstract{
Matrix models are an important class of systems in string theory and theoretical physics, with applications to random matrix theory, quantum chaos, and black holes. Hamiltonian Monte Carlo simulations and gauge/gravity duality have been used to study these systems at thermal equilibrium, and the bootstrap program has been used to efficiently determine operator expectation values by imposing positivity constraints. However, simulating real-time, non-equilibrium dynamics remains a fundamental challenge. In this work, we present the first digital quantum simulation of a bosonic matrix model, executed on the Quantinuum System Model H2 trapped-ion quantum computer. We focus on an $\mathrm{SU}(2)$ gauge theory with a quartic potential as it is simple enough to validate against exact classical solutions and yet complex enough to reflect the non-local structure of larger theories. Using the Loschmidt echo as our primary dynamical observable, we systematically decompose simulation errors into three distinct sources: Hilbert space truncation, Trotterization, and hardware noise. We demonstrate a new post-selection scheme that detects and discards gauge-symmetry violations in the Fock basis and show that at small scales it, along with zero-noise extrapolation, can give modest improvements in fidelity. These approaches struggle to scale to larger system sizes in their current implementations, emphasizing the need to move beyond them and to focus on depth reduction through improved compilation and unitary synthesis, and run-time error handling such as additional error suppression, error detection, as well as error correction approaches. This work establishes a foundation for extending digital quantum simulation to more complex matrix models---revealing that fundamental challenges in qubit resources and circuit depth remain formidable obstacles for scaling to holographically interesting regimes.
}

\maketitle

\section{Introduction}

Digital quantum simulation has become a central application driver for NISQ quantum processors~\cite{Preskill:2018jim, Daley:2022eja}, with experimental platforms now supporting dozens of high-quality qubits and native any-to-any connectivity in trapped-ion architectures~\cite{Moses:2023ozv,Ransford:2025ksn}. 
Such connectivity is particularly attractive for nonlocal Hamiltonians, where interaction graphs are dense and compilation overhead can dominate the resource budget. 
Recent trapped-ion benchmarking studies underscore both the opportunities and the practical constraints of this regime: present-day devices are often noise-limited, making algorithmic choices, including qubit encoding, circuit compilation, error suppression, and error mitigation schemes, critical for the success of the applications~\cite{Granet:2025yys, Montanez-Barrera:2025nvb, Kikuchi:2023qbb, Hayata:2026rmv, Teng:2025jva, Turro:2024shh, Dasu:2026dwm}.

In this work, we utilize QCCD-architecture trapped-ion quantum computers, specifically, the Quantinuum System Model H2 device, to benchmark the real-time digital quantum simulation of matrix models. 
Matrix quantum mechanics provides a qualitatively different stress test than widely-used benchmarks based on spin systems such as the transverse-field Ising model~\cite{Sawaya:2023nmv, kim2023evidence, Granet:2025vpn, Haghshenas:2025euj}: the Hamiltonians are typically nonlocal, involve $\mathcal{O}(N^2)$ bosonic degrees of freedom with infinite-dimensional Hilbert spaces, and—at large $N$—connect directly to problems in strongly coupled dynamics and quantum gravity. 
In particular, the Banks--Fischler--Shenker--Susskind (BFSS) matrix model~\cite{banks1997m} and its massive deformation, the Berenstein--Maldacena--Nastase (BMN) model~\cite{berenstein2002strings}, define the gauge-theory side of holographic dualities with known supergravity duals, and provide effective descriptions of D-brane dynamics in flat space (BFSS) or a pp-wave background (BMN). 
These systems therefore provide a well-motivated target class in which quantum simulation can probe dynamical observables that are difficult to access with classical tools and perturbative techniques.

A fundamental bottleneck for digital simulation of bosonic matrix models is the need to regularize the local Hilbert space by truncation. 
Any qubit-based representation of bosons (occupation-number cutoffs, digitizations in field space, or qudit-to-qubit encodings) introduces a truncation error that must be balanced against circuit depth, compilation overhead, and device noise.
Recent analyses of truncation effects in bosonic systems clarify when low-energy observables become insensitive to the cutoff and provide practical guidance for choosing truncation scales in quantum simulations~\cite{Rinaldi:2021jbg, Hanada:2022pps, Halimeh:2024bth, Brehm:2024ocy}. 
Motivated by these considerations, we focus on benchmarking the interplay among three sources of error -- Hilbert-space truncation, Trotterization, and hardware noise -- in a controlled setting where high-quality reference data is available.

Our initial benchmark target is a simplified, analytically tractable matrix quantum mechanics: a single-matrix bosonic model with quartic interaction. 
This model is solvable, not only at large $N$, but also at finite $N$ (e.g., via mappings to an effective free-fermion description for suitable sectors and observables), making it an ideal testbed for benchmarks in which discrepancies can be unambiguously attributed to the source of error or approximation:  Hilbert space truncation, digital simulation error, or experimental noise. 
Importantly, the same ingredients—bosonic truncation, nonlocal operator structure, and dense couplings—are precisely those that dominate resource requirements in more realistic holographic matrix models. 
In this sense, a solvable one-matrix model lets us benchmark techniques intended to scale to BFSS/BMN-like dynamics as hardware capabilities improve.

Digital quantum simulation of matrix models naturally complements existing classical approaches for studying matrix models. 
Lattice Monte Carlo and related Markov-chain methods have achieved substantial success in computing Euclidean (i.e., thermal equilibrium) properties of BFSS-type theories, including quantitative tests of gauge/gravity duality in regimes accessible to simulation~\cite{Pateloudis:2022ijr, Bergner:2021goh, Berkowitz:2016jlq}.
However, Euclidean Monte Carlo cannot be used to probe real-time dynamics and out-of-equilibrium observables due to the absence of a suitable analytic continuation scheme and the presence of a sign problem. The positivity/bootstrap program~\cite{lin2020bootstraps} has emerged as a powerful alternative route to nonperturbative information: starting from loop (Schwinger--Dyson) constraints and enforcing positivity of moment matrices, one can sharply bound correlators and spectra directly in the planar/large-$N$ limit for multi-matrix integrals.
This methodology was extended to matrix quantum mechanics~\cite{han2020bootstrapping}, where time-translation symmetry and gauge invariance generate constraint systems whose consistency can again be bounded using positivity, enabling precise determination of spectra and expectation values for single- and two-matrix models.
Subsequent work has broadened the range of models and critical phenomena accessible to positivity-based bootstrap techniques, strengthening the overall classical baseline against which quantum simulators can be validated (see, for instance, Refs.~\cite{kazakov2022analytic, poland2022snowmass, lin2020bootstraps, berenstein2023semidefinite, rychkov2024new} and references therein).

Against this backdrop, the quantum-simulation approach is most compelling in regimes where real-time observables are primary targets, classical Euclidean methods face severe sign/analytic-continuation barriers, and the operator structure is naturally compatible with the native connectivity of the hardware (as in trapped-ion devices). 
Matrix models sit precisely at this intersection: they are nonlocal yet structured, and they invite dynamical questions on topics such as scrambling, quenches, transport, chaos, and out-of-equilibrium correlators that are notoriously difficult to address at strong coupling. 

The main objective of this work is to evaluate the fidelity and scalability of digital quantum simulations of matrix models on trapped-ion devices using a truncated Hamiltonian representation. 
Concretely, we will:
(i) Construct qubit encodings for the truncated one-matrix quartic model and synthesize its nonlocal Hamiltonian in terms of Pauli operators. This allows us to evaluate the error introduced by the required truncation of the Hilbert space;
(ii) Implement and benchmark real-time evolution circuits (e.g., product-formula/Trotter and compilation-aware variants) and quantify algorithmic error;
(iii) Compare experimental hardware results against exact analytic results and high-precision classical numerics, quantifying the contribution of device noise to the overall error budget.

Beyond providing an application-oriented benchmark, this work yields technical insights into the many tasks required to translate the dynamics of matrix models into executable quantum programs---including the compilation and scheduling for nonlocal bosonic Hamiltonians on all-to-all-connected hardware,  principled truncation cutoff selection and error budgeting for bosonic encodings.
More broadly, this work clarifies the fundamental tension between \emph{discretization/qubitization} errors---arising from coarser truncation, larger Trotter steps, and dropped Hamiltonian terms, which keep circuits shallow but compromise physical fidelity---and \emph{depth/hardware} errors, where reducing those approximations recovers the physics but deepens the circuit and amplifies hardware noise.
Our longer-term ambition is to establish a validated workflow that connects matrix-model physics—already central in high-energy theory and gauge/gravity duality—to the experimental capabilities of trapped-ion quantum processors.

\section{Single matrix quantum mechanics \label{sec:model}}
We study a minimal matrix quantum mechanics model describing a single $N \times N$ traceless Hermitian matrix $X$. This system is the simplest representative of a rich class of matrix models. It shares the key challenges posed by matrix-valued degrees of freedom and quartic interactions with more complex theories such as the BFSS and BMN models, while remaining analytically tractable.

\subsection{Model definition}
The Hamiltonian is
\begin{equation}
    \label{eq:Hamiltonian}
    H = \text{Tr} \left( P^2 + \textrm{m}^2 X^2 + \frac{\lambda}{4N} X^4 \right) \,,
\end{equation}
where $P = D_t X \equiv \partial_t X - i[A_t, X]$ is the gauge-covariant conjugate momentum, $\textrm{m}$ is a mass constant, and $\lambda$ is the 't Hooft coupling. We gauge the global $\mathrm{SU}(N)$ symmetry, enforcing invariance under transformations $X(t) \rightarrow U^{\dagger}(t) X(t) U(t)$ for $U(t) \in \mathrm{SU}(N)$: compared to quantum field theories with spacetime degrees of freedom, in matrix models the gauge-covariant derivative acts only in time, the field strength two-form vanishes $F=0$, and hence does not appear in the Hamiltonian. There is a single dimensionless coupling, $\lambda^{1/3}/\textrm{m}$. In subsequent sections, we set $\textrm{m}=1$ and implicitly measure time in units of $1/\textrm{m}$. It will be useful to decompose $X$ in terms of the $\mathrm{SU}(N)$ generators as $X = \sum_a X_a T_a$, and similarly $P = \sum_a P_a T_a$, where $X_a$ and $P_a$ are operator-valued coefficients satisfying the canonical commutation relations ${[X_a, P_b] = i \, \delta_{ab}}$. Here, $a, b = 1, \ldots, N^2 - 1$, and the generators are normalized as $\text{Tr}(T_a T_b) = \frac{1}{2} \delta_{ab}$. The Hamiltonian becomes
\begin{equation}
    \label{eq:hamiltonian_oscillator}
    H = \frac{1}{2} \sum_a \left( P_a^2 + \textrm{m}^2 X_a^2 \right) + \frac{\lambda}{16 N^2} \left( \sum_a X_a^2 \right)^2 + \frac{\lambda}{32 N} \sum_{a,b,c,d,e} d_{abe} d_{cde} X_a X_b X_c X_d \,,
\end{equation}
where $d_{abc}$ is the symmetric tensor defined via ${ \{T_a, T_b\} = \frac{1}{2} \delta_{ab} \, \mathds{1} + \sum_c d_{abc} \, T_c}$. The gauge generators are
\begin{equation}
    \label{eq:gaugegenerators}
    G_a = i \sum_{b,c} f_{abc} \, a_b^\dagger a_c \,,
\end{equation}
where $f_{abc}$ are the structure constants of the gauge group, defined by $[T_a, T_b] = i f_{abc} T_c$. Physical states are gauge singlets, satisfying $G_a \ket{\psi} = 0$ for all $a$.

Having defined the model, we now address its tractability. Matrix models with a single matrix and a standard kinetic term admit an analytic treatment based on fermionization~\cite{anninos2020notes} (see in particular Ch.~6). The matrix degrees of freedom can be decomposed into eigenvalue and basis-change degrees of freedom. The latter are pure-gauge, and may be integrated out. The former are physical, and after taking into account the anti-symmetry of the wavefunction, may be treated as fermions. The fermions do not interact with each other, but are subjected to a potential. General eigenstates may be constructed by solving for the spectra of the individual fermions and then forming the Slater determinant to anti-symmetrize the wavefunction. Moreover, in the large-$N$ limit, collective field theory and saddle-point methods provide analytical expressions for certain quantities, such as the ground state energy~\cite{brezin1978planar}.

For $N=2$, the system admits an even simpler description that we exploit for benchmarking. The symmetric tensor vanishes, $d_{abc} = 0$, and the potential in Eq.~\eqref{eq:hamiltonian_oscillator} simplifies considerably. The local isomorphism between $\mathrm{SU}(2)$ and $\mathrm{SO}(3)$ allows the three matrix degrees of freedom to be interpreted as the coordinates of a particle in a central potential. Additionally, gauge invariance restricts the dynamics to the $\ell=0$ sector, reducing the Schr\"odinger equation to a one-dimensional radial problem. Thus, the single-matrix model is sometimes described as ``solvable'', though this should be understood loosely: the reduced problem is still that of an anharmonic oscillator, which does not admit a closed-form solution.

\subsection{Radial reduction for \texorpdfstring{$N=2$}{N=2}}
\label{sec:exact_solution}
The matrix $X$ is a traceless Hermitian $2 \times 2$ matrix transforming in the adjoint representation of $\mathrm{SU}(2)$. Expanding $X = \sum_{a=1}^3 X_a \sigma_a/2$ in the Pauli basis gives three Hermitian operator degrees of freedom $X_a$, corresponding to three coupled quantum oscillators. 
In the position representation where $X_a = x_a$ and $P_a = -i \partial_{x_a}$, the time-independent Schr\"odinger equation is:
\begin{equation}
    \left( - \frac{1}{2}\nabla^2 + V(x) \right) \psi(x) = E \, \psi(x) \,,
\end{equation}
where 
\begin{equation}
    V(x) = \frac{\textrm{m}^2}{2} \sum_{a=1}^{3} x_a^2 + \frac{\lambda}{64} \left( \sum_{a=1}^{3} x_a^2 \right)^2 \,.    
\end{equation}

The potential is spherically symmetric, which suggests mapping from Cartesian coordinates $(x_1, x_2, x_3) \in \mathbb{R}^3$ to spherical coordinates $(r, \Omega)$, where $r = \sqrt{\sum_a x_a^2}$ is the radius and $\Omega$ denotes the angular coordinates of the 2-sphere. The potential becomes simply
\begin{equation}
    V(r) = \frac{\textrm{m}^2}{2} r^2 + \frac{\lambda}{64} r^4 \,.
\end{equation}

Naturally, the wavefunction can be decomposed in terms of spherical harmonics, that is, we assume ${\psi(r, \Omega) = R(r) Y_{\ell m}(\Omega)}$, for $R$ the radial wavefunction, and $Y_{\ell, m}$ the usual spherical harmonics. 
Introducing the rescaled variable ${u(r) = r \, R(r) }$ results in the standard radial Schr\"odinger equation: 
\begin{equation}\label{eq:radial_equation}
    \left( - \frac{1}{2} \frac{\dd^2}{\dd r^2} + \frac{\ell (\ell + 1)}{2r^2} + V(r) \right) u(r) = E \, u(r) \,.
\end{equation}

The gauge symmetry has not yet been imposed. 
For $\mathrm{SU}(2)$, the structure constants are given by the completely anti-symmetric tensor, $f_{abc} = \varepsilon_{abc}$, and the gauge generators Eq.~\eqref{eq:gaugegenerators} coincide with the angular momentum generators, $G_a = \varepsilon_{abc} \, x_b \, p_c$. Physical states therefore correspond to spherically symmetric states with $\ell = m = 0$.

Additionally, square-integrability requires the boundary conditions $u(r=0) = 0$ and $\lim_{r\rightarrow \infty} u(r) = 0$. In the non-interacting case, $\lambda = 0$, this reduces to the equation of a 1D simple harmonic oscillator, restricted to odd eigenstates due to the boundary condition at $r=0$. The free spectrum is therefore
\begin{equation}
    \label{eq:free_oscillator_spectrum}
    E_n(\lambda = 0) = \textrm{m} \left( 2 n + \frac{3}{2} \right) \,, \quad n \in \mathbb{N}_0.
\end{equation}
The case with $\lambda \neq 0$ may be treated either numerically or through a perturbative treatment. In App.~\ref{app:numerical_solution}, we pursue a numerical approach based on spectral collocation methods.

\subsection{The Loschmidt echo \label{sec:loschmidt}}
For our benchmarking experiments, we focus on the Loschmidt echo as the primary observable. For a reference state $\ket{\varphi}$, the Loschmidt echo is defined as~\cite{gorin2006dynamics}:
\begin{equation}\label{eq:loschmidt_echo}
    M_{\varphi}(t) = \left| \bra{\varphi} e^{i H_0 t} e^{-i H t} \ket{\varphi} \right|^2 \,,
\end{equation}
where $H_0$ is the free, non-interacting Hamiltonian. Physically, $M_{\varphi}(t)$ quantifies the fidelity between a state evolved under the full interacting dynamics and one evolved under only the free Hamiltonian, providing a sensitive probe of interaction effects~\cite{yan2020information}. In this work, we will take the reference state to be the free vacuum $\ket{\mathrm{vac}}$.

As a benchmarking diagnostic, the Loschmidt echo offers several advantages: it provides a single scalar quantity that captures the cumulative effects of Trotter error, truncation error, and hardware noise, while its Fourier transform yields spectral information about the interacting Hamiltonian. To demonstrate this last point, we decompose the free vacuum in terms of energy eigenstates of the \textit{interacting} system $\psi_n$: $\ket{\mathrm{vac}} = \sum_{n} c_n \ket{\psi_n}$. The forward time evolution is then straightforward; each eigenstate acquires a phase $e^{-i E_n t}$. The backward evolution is trivial as $\ket{\mathrm{vac}}$ is an eigenstate of $H_0$ and the term $e^{-i H_0 t}$ therefore contributes a global phase that drops out once the absolute value is taken. The echo is then
\begin{equation}
    M_{\mathrm{vac}}(t) = \left| \sum_{n=0}^{\infty} \left| c_n \right|^2 e^{-i E_n t} \right|^2  \,.
\end{equation}

\begin{figure}[H]
\centering
\includegraphics[width=\linewidth]{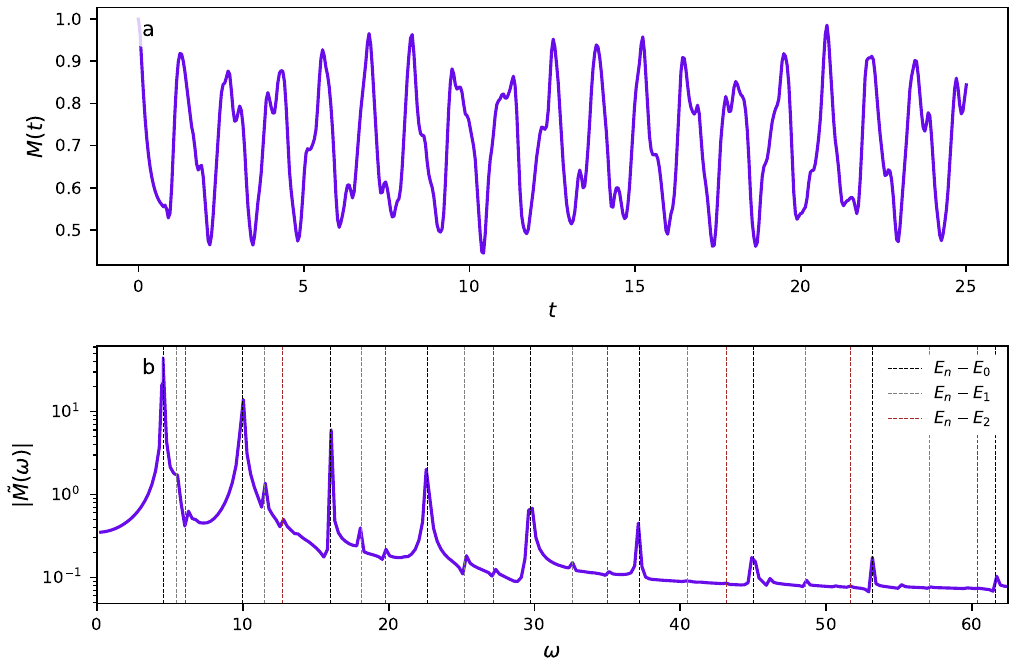}
\caption{
The Loschmidt echo and Fourier transform. (a) The Loschmidt echo as a function of time for $\lambda = 40$. (b) The absolute value of the Fourier-transformed echo, $|\tilde{M}(\omega)|$, as a function of frequency. Vertical dashed lines correspond to selected energy gaps $E_{n_2} - E_{n_1}$ to guide the eye.
}
\label{fig:Loschmidt_echo_and_FFT}
\end{figure}

The echo encodes spectral information that can, in principle, be extracted from experimental data. Over short times, the echo decays quadratically as ${M_{\mathrm{vac}}(t) = 1 - \sigma_E^2 \, t^2 + \mathcal{O}(t^4)}$, where $\sigma_E^2 = \langle H^2 \rangle_{\mathrm{vac}} - \langle H \rangle_{\mathrm{vac}}^2 $ is the energy variance in the free vacuum. Furthermore, the sum over eigenstates shows that $M_\mathrm{vac}(t)$ oscillates at frequencies corresponding to energy differences ${\omega_{n_1 n_2} := E_{n_2} - E_{n_1}}$. The Fourier-transformed echo is
\begin{equation}
    \label{eq:fourier_transform_echo}
    \tilde{M}_{\mathrm{vac}}(\omega) = \int_{-\infty}^{\infty} \dd t \, e^{i \omega t} M_{\mathrm{vac}}(t) = 2\pi \sum_{n_1, n_2} |c_{n_1}|^2 |c_{n_2}|^2 \, \delta\left(\omega - \omega_{n_1 n_2} \right) \,.
\end{equation}
The peak positions give the energy differences, while the peak weights give products of occupation probabilities. This is illustrated in Fig.~\ref{fig:Loschmidt_echo_and_FFT}. Panel (a) shows the Loschmidt echo $M(t)$, and Panel (b) shows the magnitude of its Fourier transform $|\tilde{M}(\omega)|$. The vertical dashed lines indicate the spectral gaps $\omega_{n_1 n_2}$ for $n_1 = 0, 1, 2$. Peaks corresponding to transitions from the ground state are clearly visible.
The finite peak width is set by the observation time, $\Delta \omega \sim 2\pi/T$. Both the low-lying energies $E_n$ and gaps $\omega_{n_1 n_2}$ grow monotonically with $\lambda$ (see Fig.~\ref{fig:energy_and_gap_vs_coupling} in App.~\ref{app:numerical_solution}). Consequently the smallest gap satisfies $\omega_{01}\ge 2m$, so the longest intrinsic timescale is bounded by $T_{\max}=2\pi/\omega_{01}\le \pi/m$, setting the minimum time resolution required to extract spectral features from device data.

The Loschmidt echo $M_\mathrm{vac}(t)$ is closely related to the spectral form factor $K(t) = |\sum_n e^{-iE_n t}|^2$, a standard diagnostic for quantum chaos. Our expression differs only in the weighting: $|c_n|^2$ from the initial state overlap versus uniform or Boltzmann weights. The Fourier transform of the spectral form factor yields the two-point spectral correlation function, whose structure distinguishes Poisson statistics (integrable) from Wigner-Dyson statistics (chaotic)—in particular, the characteristic ``dip-ramp-plateau'' of $K(t)$ is a universal signature of level repulsion in chaotic systems \cite{cipolloni2023spectral}. Exploring this connection may provide a more principled framework for interpreting our benchmarking data.

\section{Digital quantum simulation protocol}
In this section, we explain how the dynamics of the single matrix model can be simulated using a qubit-based digital quantum computer.

\subsection{Fock-space truncation and qubit encoding \label{sec:truncation}}
The Hilbert space of a traceless $N\times N$ Hermitian matrix is infinite-dimensional, so any qubit implementation requires a cutoff. We adopt the mode-by-mode Fock-space truncation of Ref.~\cite{Gharibyan:2020bab}. Writing $X=\sum_a X_a T_a$ yields $N^2-1$ harmonic-oscillator modes $X_a$; for each mode $a$ we retain only the lowest $\Lambda$ number states $\{\ket{0}_a,\ket{1}_a,\ldots,\ket{\Lambda-1}_a\}$---corresponding to the lowest-energy states of the free Hamiltonian $H_0 = m \sum_{a=1}^{N^2-1} \left(N_a + \frac{1}{2} \right)$, where $N_a$ is the oscillator $a$ number operator. The resulting truncated Hilbert space is
\begin{equation}
    \mathcal{H}_{\mathrm{trunc}}
    = \bigotimes_{a=1}^{N^2-1} \mathrm{span}\{\ket{0}_a,\ket{1}_a,\ldots,\ket{\Lambda-1}_a\}\,.
\end{equation}
We choose $\Lambda=2^K$ so that each oscillator is encoded in $K$ qubits, giving a total number of qubits
\begin{equation}
    n_Q = (N^2-1)K\,.
\end{equation}

The truncation scheme is described in detail in App.~\ref{app:truncation}. Operationally, an operator $O$ is mapped to a truncated operator $O^{(\Lambda)}$ acting on $\mathcal{H}_{\mathrm{trunc}}$. In particular, after choosing a binary encoding of the truncated Fock basis on $n_Q$ qubits, every $O^{(\Lambda)}$ admits a Pauli-string expansion,
\begin{equation}
    \label{eq:operator_pauli_decomposition}
    O^{(\Lambda)} = \sum_{P \in \mathcal{P}_{n_Q}} c_P \, P \,,
\end{equation}
where $\mathcal{P}_{n_Q}$ is the $n_Q$-qubit Pauli group, $P$ are Pauli strings, and $c_P$ are complex coefficients. We will use this representation to compile Trotterized time-evolution circuits and to evaluate observables.

An important consequence of the cutoff is that gauge symmetry is only approximate at finite $\Lambda$: the truncated generators $G_a^{(\Lambda)}$ no longer close the $\mathrm{SU}(N)$ algebra and, in general, $[G_a^{(\Lambda)},H^{(\Lambda)}]\neq 0$. These violations are dominated by boundary effects near the Hilbert space cutoff (i.e., states with support on the highest occupation numbers) and are exponentially suppressed for
low-lying states at moderate $\Lambda$. For example, the ground-state Casimir expectation value was found to decay exponentially with the cutoff in Ref.~\cite{Rinaldi:2021jbg} using exact diagonalization techniques, and in Ref.~\cite{Brehm:2024ocy} using tensor-network techniques\footnote{See Ref.~\cite{Hanada:2025goy} for more details on the effects of gauge symmetry in quantum simulations}.

\subsection{Gauge singlets in the Fock basis for \texorpdfstring{$N=2$}{N=2} \label{sec:gaugesinglet}}
Having shown in Sec.~\ref{sec:exact_solution} that $\mathrm{SU}(2)$ gauge singlets coincide with the spherically symmetric sector $\ell=m=0$, we now describe their representation in the Fock basis $\ket{n_x n_y n_z}$. The discussion below is exact prior to truncation; cutoff effects are discussed at the end of this section.

The total number operator, $N_{\mathrm{tot}} = \sum_a N_a$, commutes with the gauge generators, $[G_a, N_{\mathrm{tot}}] = 0$, which means that we may simultaneously consider states of definite gauge charge and total particle number (though note that in the interacting case, we cannot also simultaneously diagonalize the Hamiltonian). In the $\mathrm{SU}(2)$ case (with modes $a\in\{x,y,z\}$), this implies an orthonormal singlet basis $\{\ket{2k}_{\mathrm{singlet}}\}_{k\ge 0}$ at even total occupation, which can be generated from the free vacuum $\ket{\mathrm{vac}}$ by repeated action of the singlet pair-creation operator $\vec a^{\dagger}\!\cdot\!\vec a^{\dagger}=(a_x^\dagger)^2+(a_y^\dagger)^2+(a_z^\dagger)^2$:
\begin{equation}
    \label{eq:gauge_singlet_fock_expansion}
    \ket{2k}_{\mathrm{singlet}} = \frac{1}{\mathcal N_k}\bigl(\vec a^{\dagger}\!\cdot\!\vec a^{\dagger}\bigr)^k\ket{\mathrm{vac}}\,,
\end{equation}
with $\mathcal N_k$ a normalization constant. These states diagonalize the free Hamiltonian but not the interacting one, since the quartic term mixes different $N_{\mathrm{tot}}$ sectors. Consequently, the interacting singlet energy eigenstates admit the expansion
\begin{equation}
    \label{eq:radial_ket_gauge_expansion}
    \ket{n,\ell=0,m=0}=\sum_{k=0}^{\infty}\alpha^{(n)}_{k}(\lambda)\,\ket{2k}_{\mathrm{singlet}}\,,
    \end{equation}
where $n=0,1,2,\ldots$ labels the eigenstates in increasing energy.

The relation Eq.~\eqref{eq:gauge_singlet_fock_expansion} allows the dimension of the gauge singlet sector retained at a given truncation level to be computed. For $\Lambda=2^K$, each mode supports occupations $n_a\in\{0,1,\ldots,2^K-1\}$, so the largest even occupation is $2^K-2$. Since $\ket{2k}_{\mathrm{singlet}}$ has total occupation $2k$, the truncation implies $k_{\text{max}} = 2^{K-1}-1$. 
Hence the truncated singlet sector contains the basis states $\ket{2k}_{\mathrm{singlet}}$ with $k=0,1,\ldots,k_{\text{max}}$, and therefore has dimension $k_{\text{max}}+1=2^{K-1}$. The full truncated Hilbert space has dimension $2^{n_Q}=2^{3K}$, so the vast majority of truncated states lie outside the singlet sector and carry nonzero gauge charge.

We now address the effects of truncation. As discussed above, the truncated gauge generators $G_a^{(\Lambda)}$ no longer satisfy the exact $\mathrm{SU}(2)$ algebra, and $[G_a^{(\Lambda)}, H^{(\Lambda)}] \neq 0$. In particular, $[G_a^{(\Lambda)}, \vec{a}^\dagger \cdot \vec{a}^\dagger] \neq 0$, so the states constructed via Eq.~\eqref{eq:gauge_singlet_fock_expansion} acquire small gauge charge from boundary terms near the truncation edge, and eigenstates of $H^{(\Lambda)}$ are not exact gauge singlets. However, these violations are exponentially suppressed for low-lying states at moderate $\Lambda$, ensuring that the post-selection scheme described above remains effective in practice.

\subsection{Trotterization and compilation \label{sec:circuit}}
\begin{figure}[H]
\centering
\includegraphics[width=\linewidth]{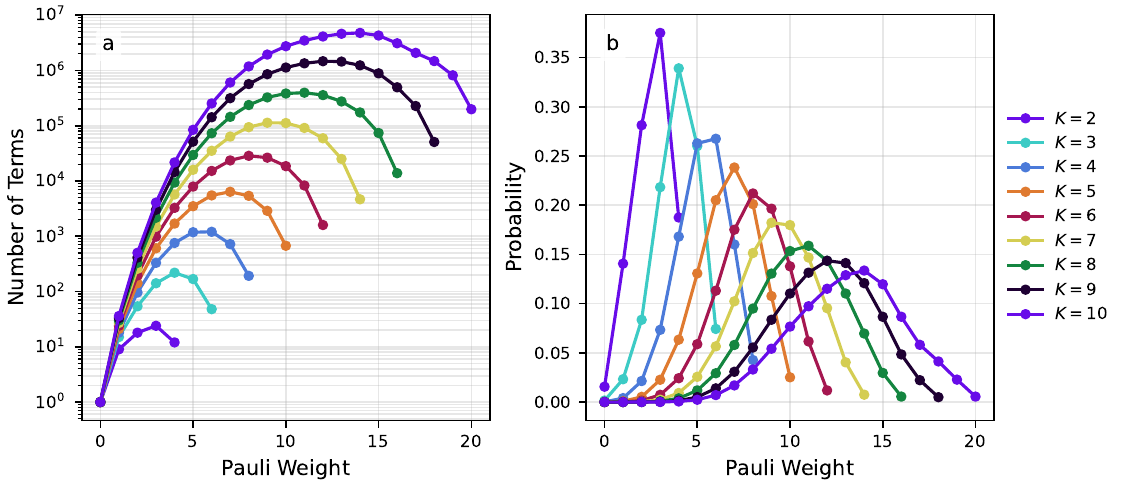}
\caption{
Pauli decomposition of the truncated Hamiltonian.
(a) The number of Pauli terms as a function of the Pauli weight.
(b) The normalized probability distribution of Pauli weights.
}
\label{fig:Hamiltonian_pauli_decomposition}
\end{figure}

The truncation scheme described in Sec.~\ref{sec:truncation} results in a Hamiltonian that takes the form of a weighted sum of Pauli strings, $H^{(\Lambda)} = \sum_{P \in \mathcal{P}_{n_Q}} c_P P$, c.f. Eq.~\eqref{eq:operator_pauli_decomposition}. To characterize the complexity of this encoding, we analyze the resulting operator structure in Fig.~\ref{fig:Hamiltonian_pauli_decomposition}. Figure~\ref{fig:Hamiltonian_pauli_decomposition}(a) displays the total number of terms in the Pauli expansion as a function of their weight (the number of qubits on which the operator acts non-trivially) for $K$ (bits per oscillator) ranging from 2 to 10. The total number of terms scales roughly as $\sim 5^K$, with a maximum Pauli weight of $2K$. While exponential, this growth is significantly favorable compared to the theoretical bound of $4^{3K} = 64^K$ corresponding to a operator with support over all $3K$-qubit Pauli strings, confirming that the Hamiltonian occupies an exponentially sparse sector of the full operator space. In Fig.~\ref{fig:Hamiltonian_pauli_decomposition}(b), we present the normalized weight distribution. As $K$ increases, the support of the distribution shifts continuously toward higher weights, indicating that the encoding becomes increasingly non-local with $K$. 
 
We approximate the time evolution operator $e^{-i H t}$ with the first-order Lie-Trotter formula applied to the truncated Hamiltonian. The Trotterized time evolution operator for $r$ steps and time-step $\dd t = t / r$ is
\begin{equation}
U(t; r) = \left( \prod_{P \in P_{n_Q}} e^{-i c_P P \dd t} \right)^{r} \,.
\end{equation}
This approximation would be exact if all Pauli terms mutually commute, or in the limit $r \to \infty$. The per-step error is controlled by the commutators between terms: 
$\mathcal{O}\!\left( \dd t^2 \sum_{P < P'} |c_P c_{P'}| \, \norm{[P, P']} \right)$,
where the sum runs over distinct pairs of non-commuting Pauli operators, and $\norm{ \cdot }$ is the spectral or infinity norm. The error can also be seen to scale as $\mathcal{O}(\lambda^2 \dd t^2)$---the non-interacting Hamiltonian $H_0$ is diagonal in the $Z$-basis, and ${[H_0, H_{\mathrm{int}}] \propto \lambda}$. 

The Lie–Trotter product must then be compiled into a quantum circuit. We leverage the freedom to reorder Pauli terms to group those acting on disjoint qubit subsets, enabling parallel execution and reducing circuit depth. For the Loschmidt echo, an additional simplification applies: when the reference state $\ket{\varphi}$ is the free vacuum, the free evolution $e^{iH_0 t}$ contributes only a global phase (since $H_0$ is diagonal in the $Z$ basis), and can be omitted. Likewise, using the first-order splitting $e^{-iHt}\approx e^{-iH_{\mathrm{int}}t}e^{-iH_0 t}$, the final $e^{-iH_0 t}$ factor can be dropped for the same reason, so only the interacting evolution needs to be implemented.

As the truncation level $K$ increases, the rapid growth in the number of terms in the Pauli expansion of $H^{(\Lambda)}$ leads to a corresponding increase in Trotter circuit complexity. Table~\ref{tab:circuit_metrics} illustrates the scaling of the single-step Trotter circuit depth with respect to $K$; notably, these metrics are independent of the coupling $\lambda$ for generic values. The total depth and gate count for an $r$-step Trotter evolution scale linearly with the single-step metrics, imposing two practical limitations. First, circuits with two-qubit (2Q) count $\gtrsim 1000$ are effectively noise-dominated on the Quantinuum H2-2 device for our choice of a global observable. Second, device credits---a scarce resource---scale with circuit depth. To ensure a meaningful signal within our credit budget, we restricted hardware experiments to the $K=2$ case.

\begin{table}[h]
\caption{Complexity of a single Trotter step circuit as a function of the number of qubits per oscillator $K$.}
\label{tab:circuit_metrics}
\centering
\begin{tabular}{c c c c}
\hline
$K$ & Circuit Depth & 2Q Depth & 2Q Gate Count \\
\hline
2 & 101 & 52 & 64 \\
3 & 1131 & 604 & 942 \\
4 & 7907 & 4227 & 6661 \\
\hline
\end{tabular}
\end{table}

\subsection{Error mitigation \label{sec:mitigation}}

Hardware noise is the dominant error source in our deepest circuits. Although general error-mitigation protocols that provide rigorous accuracy guarantees exist (for instance, probabilistic error cancellation \cite{temme2017error}), these come at the cost of an exponential sampling overhead with the circuit depth \cite{takagi2022fundamental}. Our circuits are depth-limited on current devices, and we therefore consider two comparatively lightweight post-processing techniques to partially mitigate hardware noise. For the system sizes considered here, these methods remain computationally tractable; scalability to deeper or wider circuits is a separate question we do not address.

\paragraph{Zero-noise extrapolation (ZNE).}
Current quantum hardware is noisy, so we apply a low-overhead error-mitigation technique: digital zero-noise extrapolation (ZNE)~\cite{zne}. ZNE estimates the noiseless expectation value by deliberately scaling the circuit noise, measuring the observable at several effective noise levels, and extrapolating back to ``zero noise''. We implement noise scaling via two-qubit gate folding: each two-qubit gate is replaced by $U_{2Q} \;\longrightarrow\; U_{2Q} \,(U_{2Q}^\dagger U_{2Q})^k$, which preserves the ideal unitary (since $U_{2Q}^\dagger U_{2Q}=\mathds{1}$) while increasing the number of noisy two-qubit gates by a factor $f=2k+1$. In our experiments we use two scale factors, $f \in \{1,3\}$ (i.e., $k=0,1$), and perform a linear extrapolation in $f$ to estimate the zero-noise limit.

\paragraph{Gauge singlet post-selection.}
Equation~\eqref{eq:gauge_singlet_fock_expansion} shows that singlet states have support only on Fock basis states $\ket{n_x n_y n_z}$ with even occupation in each mode. This leads to a simple error-detection rule for digital time evolution initialized in a singlet: any measurement outcome with an odd occupation in any mode indicates a bit-flip error and can be discarded in post-processing. Concretely, our post-selection keeps only shots with $(n_x,n_y,n_z)$ all even (and renormalizes the retained sample), which suppresses bit-flip-induced leakage out of the singlet subspace and therefore increases the fidelity of estimated observables. Since a given mode is even with probability $1/2$, only a fraction $(1/2)^3=1/8$ of Fock states satisfy the all-even constraint. 
 
\section{Results}
We now present the results of our benchmarking experiments. As described above, digital quantum simulation of matrix models requires a sequence of approximations, each introducing distinct errors that compound in the final measurement. First, the infinite towers of states, one for each oscillator, are truncated so that only a finite set of states are retained. Next, the exact time evolution is approximated via Trotterization. Lastly, the circuit is executed and measured on noisy hardware. We analyze each error source---Hilbert space truncation, Trotterization, and device noise---in turn. We also consider error mitigation strategies to partially correct the noisy measurement results. As discussed in Sec.~\ref{sec:circuit}, we present simulated and experimental circuit execution data for the truncation level $K=2$ (6 qubits). Even at this scale, the Trotterized evolution yields circuit depths that approach the device's coherence limits.

\subsection{Truncation error}
\begin{figure}[H]
\centering
\includegraphics[width=\linewidth]{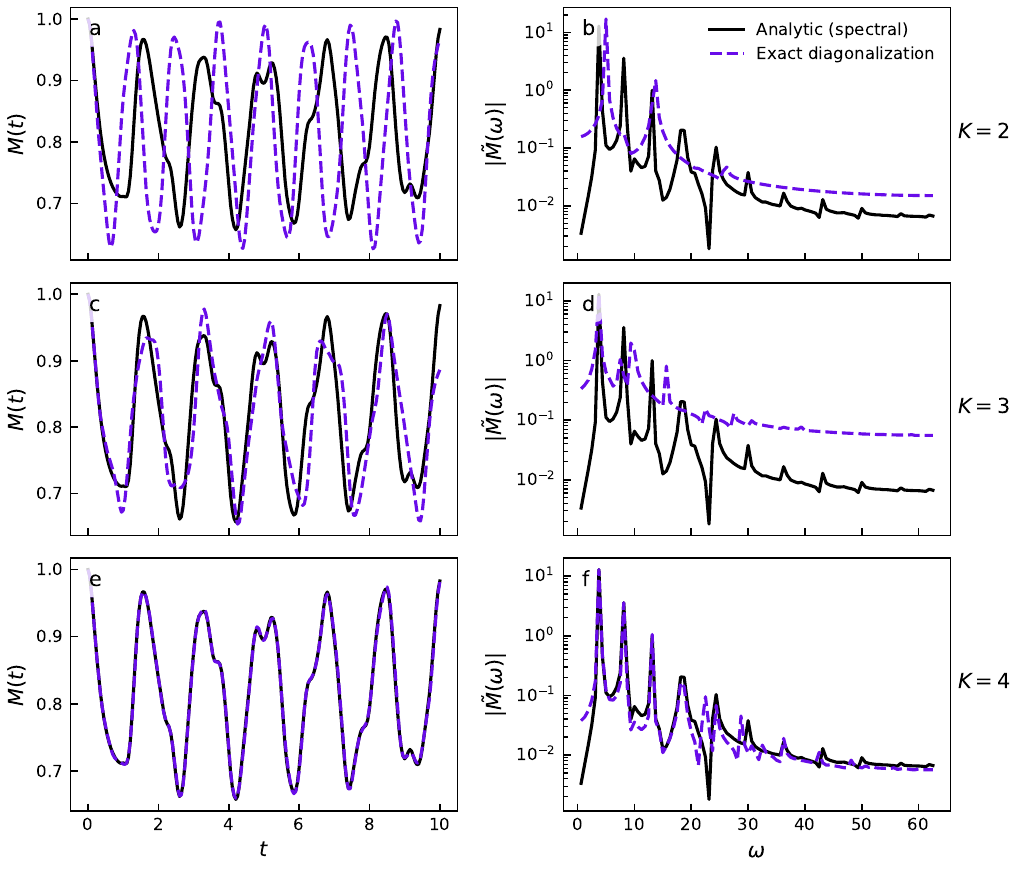}
\caption{
Effect of Hilbert space truncation on the Loschmidt echo. (a, c, e): the echo $M(t)$ as a function of time. (b, d, f): the magnitude of the Fourier transform $|\tilde{M}(\omega)|$. Each row corresponds to a different truncation level $K$ (bits per oscillator), with the exact solution, corresponding to $K \rightarrow \infty$ shown in black for comparison. The data were generated numerically via exact diagonalization for the truncated model and the spectral collocation method described in App.~\ref{app:numerical_solution} for the exact model. The data corresponds to a model with $\lambda = 20$. The total simulation time is $T=10$, and time points have been sampled with a time increment of $\dd t = 0.05$.
\label{fig:Loschmidt_echo_exact_vs_qubitized}
}
\end{figure}

Fig.~\ref{fig:Loschmidt_echo_exact_vs_qubitized} depicts the Loschmidt echo and its Fourier transform for both the truncated and non-truncated model. The Fourier transform exhibits a series of peaks, corresponding to spectral gaps, c.f., Eq.~\eqref{eq:fourier_transform_echo}. In the case of the exact, un-truncated model, the three most prominent peaks correspond to the gap between the ground state and the first excited states, i.e., to $\omega_{0n} = E_n - E_0$ for $n=1,2,3$. The truncation error rapidly decreases as $K$ increases, with $K=4$ leading to negligible error. In particular, for $K=4$, the locations of the three most prominent peaks in the Fourier transform agree to those of the exact, un-truncated model up to the frequency resolution $\Delta\omega = 2\pi/T$ set by the finite observation time. To further quantify the discrepancy, Table~\ref{tab:truncation_error} shows the time-averaged mean absolute error (MAE) for $M(t)$, as well as the amplitude error of the peaks in $\tilde{M}(\omega)$. The peak amplitudes agree to within $2\%$ for $K=4$, confirming the accuracy of this truncation level. Ref.~\cite{Rinaldi:2021jbg} investigated matrix models using the same truncation and compact qubit encoding scheme used here and observed doubly exponential convergence for select gauge-invariant observables; our results are consistent with those findings. 

\begin{table}[h]
\caption{Truncation error as a function of the Hilbert space cutoff $K$ (bits per oscillator). Time MAE (mean absolute error) is the absolute error of the Loschmidt echo, computed relative to the non-truncated value, and averaged over time. The mean peak amplitude error compares the three dominant peaks in the Fourier-transformed echo against the exact solution.}
\label{tab:truncation_error}
\centering
\begin{tabular}{c c c}
\hline
$K$ & Time MAE & Mean peak amplitude error \\
\hline
2 & $1.26 \times 10^{-1}$ & 77.7\% \\
3 & $3.57 \times 10^{-2}$ & 53.2\% \\
4 & $2.23 \times 10^{-3}$ & 2.0\% \\
\hline
\end{tabular}
\end{table}

\subsection{Trotterization error}
Trotterization involves a trade-off between accuracy and accessible evolution time. For a first-order scheme, the per-step error scales as $\mathcal{O}(\dd t^{2})$, favoring small $\dd t$, while resolving the dynamics of interest---especially the low-energy gaps visible in the Loschmidt echo---requires a long total evolution time $T=r\,\dd t$, favoring large $T$ and hence many steps $r$. However, the number of Trotter steps is constrained by both device noise and available device credits. Circuit depth scales linearly with $r$: for $K=2$, a single step uses 64 two-qubit (2Q) gates and has 2Q depth 52 (Tab.~\ref{tab:circuit_metrics}). Larger $r$ therefore increases the accumulated hardware error and requires running more (and deeper) circuits on device.

The ZNE mitigation strategy further limits the number of Trotter steps. At folding level $f=2k+1$, each 2Q gate is replaced by $f$ 2Q gates, so the 2Q count (and roughly the 2Q depth) grows by $\sim f$. This tightens the depth budget and limits us to $r\lesssim 5$ while maintaining signal across the ZNE data points. We therefore choose $\dd t=0.1$ as a compromise between Trotter error and reachable $T$ within the hardware depth budget.

\begin{figure}[H]
\centering
\includegraphics[width=\linewidth]{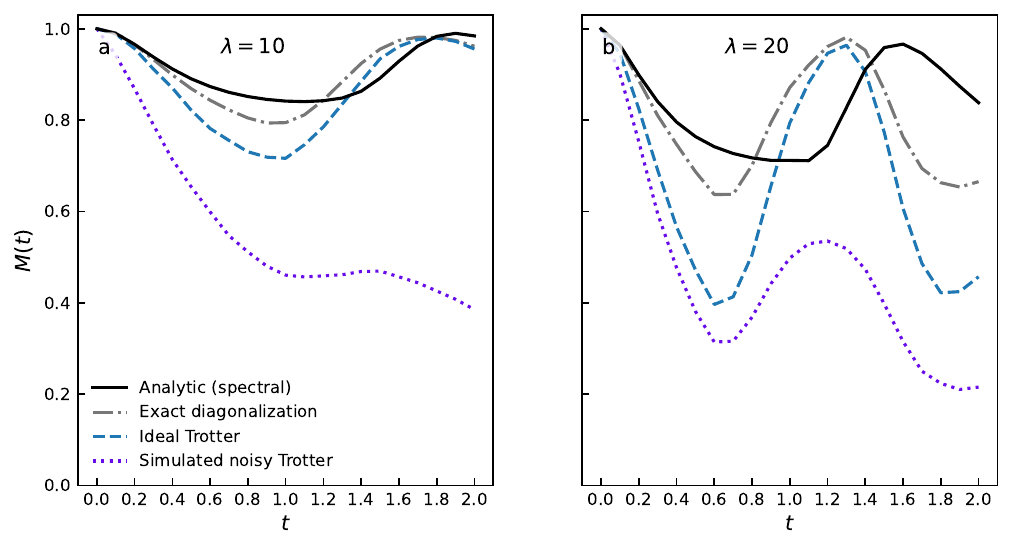}
\caption{
Loschmidt echo for truncation level $K=2$ simulated with the Qiskit Aer noise simulator. Parameters: $\dd t = 0.1$, depolarizing noise $p = 10^{-3}$ on two-qubit gates, $30{,}000$ shots.}
\label{fig:echo_aer_simulation}
\end{figure}

To isolate Trotter error in the Loschmidt echo, we compare exact diagonalization of the truncated Hamiltonian with (i) noiseless Trotterized evolution and (ii) Trotterized evolution under a depolarizing noise model calibrated to Quantinuum; for context we also show the untruncated analytic result. Figure~\ref{fig:echo_aer_simulation} summarizes these comparisons. For $\lambda=10$, the noiseless Trotter error is $0.1-9.8\%$ relative to exact diagonalization, while the noisy Trotter error increases to $4.9-60.0\%$. For $\lambda=20$, the corresponding ranges are $1.5-37.8\%$ (noiseless) and $6.5-67.9\%$ (noisy). Although the noise is simulated here, it provides a rough proxy for device performance and motivates the error-mitigation procedures applied to the hardware results in Sec.~\ref{sec:mitigation}.

\subsection{Hardware error}
\begin{figure}[H]
\centering
\includegraphics[width=\linewidth]{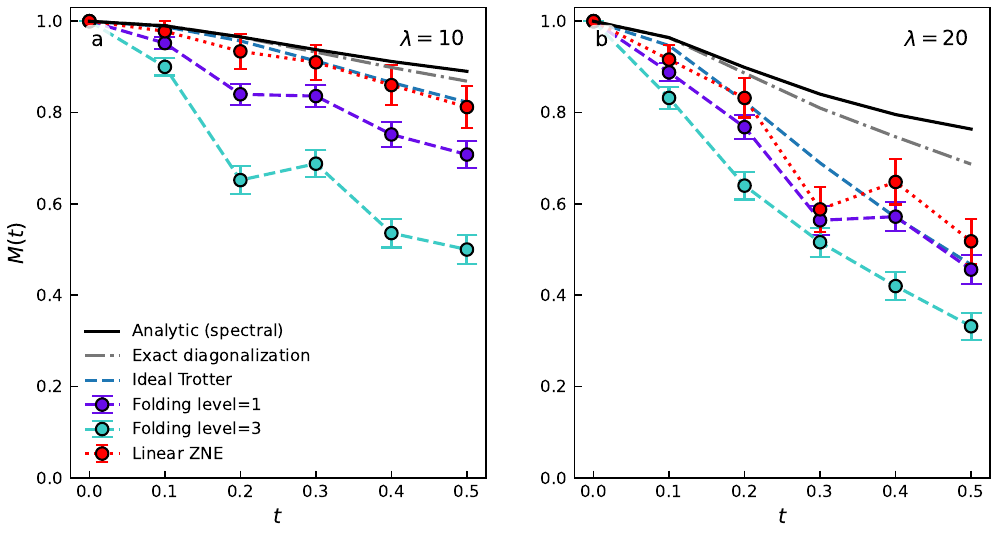}
\caption{
Loschmidt echo for truncation level $K=2$ measured on Quantinuum H2-2 device ($\dd t = 0.1$, 250 shots per point). Error bars show one standard deviation. Hardware data is shown for folding levels 1 (no folding) and 3, along with the zero-noise extrapolated (ZNE) result. Reference curves: exact solution of the Schr\"odinger equation, exact diagonalization of the truncated system, and ideal Trotterized evolution.
}
\label{fig:echo_h2-2}
\end{figure}

Having considered the truncation and Trotter errors, we next turn to the hardware results. 
Fig.~\ref{fig:echo_h2-2} depicts the digital quantum simulation of the Loschmidt echo on the Quantinuum H2-2 device. 
The raw data (folding level 1) exhibits significant deviation from the ideal Trotter result. We attempt to mitigate this using ZNE, as described in Sec.~\ref{sec:mitigation}. ZNE requires executing circuits at multiple noise levels; folding level 3 is also shown, which exhibits greater error as expected, enabling extrapolation to zero noise. For $\lambda = 10$, ZNE reduces the mean absolute error by 72\%, with improvement ranging from $72-96\%$ across time steps. For $\lambda = 20$, ZNE improves accuracy at early times ($t \leq 0.3$), reducing errors by $19-85\%$. However, at later times ($t \geq 0.4$), ZNE produces results worse than the raw data. This occurs when the raw error is already small ($10^{-3}$--$10^{-2}$), well below the shot noise floor of approximately $1/\sqrt{250} \approx 0.06$. In this regime, ZNE is extrapolating noise rather than systematic error, and can overshoot the true value. 
To wit, we note that ZNE's practical efficacy is sensitive to the noise regime. For circuits with moderate depth, where expectation values retain meaningful signal under noise scaling, ZNE can yield substantial error reduction as the results show, including the deepest circuits for the weaker coupling value of $\lambda=10$. However, as the noise level increases, folding risks suppressing the residual signal entirely, leaving insufficient dynamic range for reliable extrapolation.

\subsection{Symmetry-based benchmarking and mitigation}
\begin{figure}[H]
\centering
\includegraphics[width=\linewidth]{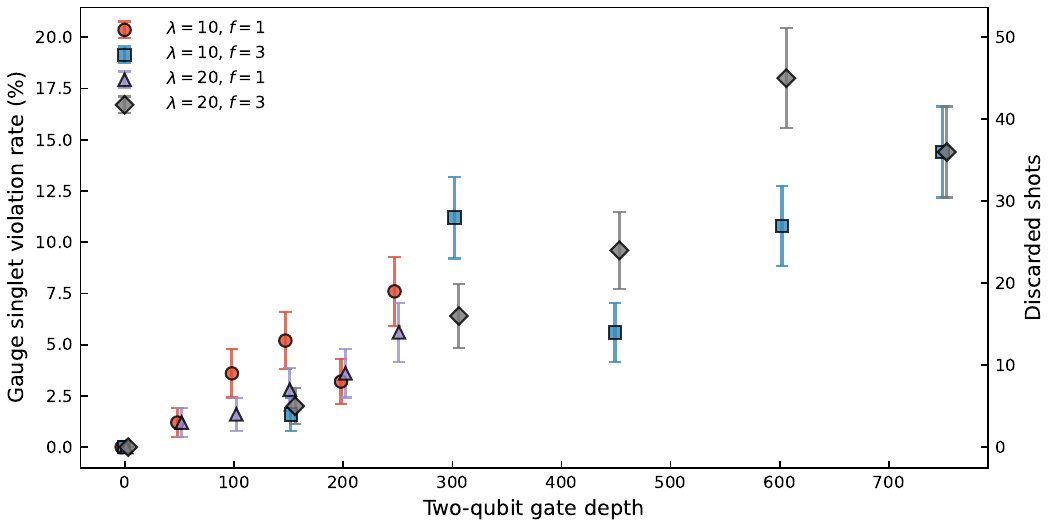}
\caption{
Measurement shots violating the gauge singlet condition as a function of 2Q depth. Error bars indicate one standard deviation. A small horizontal jitter has been added to help distinguish data points sharing the same $x$-value. Data correspond to truncation level $K=2$.}
\label{fig:gauge_violation_rate_vs_effective_depth}
\end{figure}

We can further benchmark the device by evaluating the extent to which the observed shots violate the gauge singlet constraint. Each shot yields a bitstring that encodes the occupation numbers for each oscillator, $(n_x, n_y, n_z)$. As described in Sec.~\ref{sec:gaugesinglet}, the singlet sector is restricted to all-even occupation numbers, and thus any measurement yielding one or more odd occupation numbers necessarily indicates an error. The number of errors can be expected to increase with the Trotter simulation time, or equivalently, the circuit depth. Fig.~\ref{fig:gauge_violation_rate_vs_effective_depth} shows the fraction of violating shots as a function of the 2Q depth. Data are shown for the two cases that we have hardware data for---folding levels 1 (no folding) and 3. The discard rate grows with the two-qubit gate depth of the circuit, and is stronger for the weaker coupling value considered. The maximal discard rate for the unfolded circuits is $7.6\% \pm 1.7\%$ for $\lambda=10$, and $5.6\% \pm 1.5\%$ for $\lambda=20$. 

\begin{figure}[H]
\centering
\includegraphics[width=\linewidth]{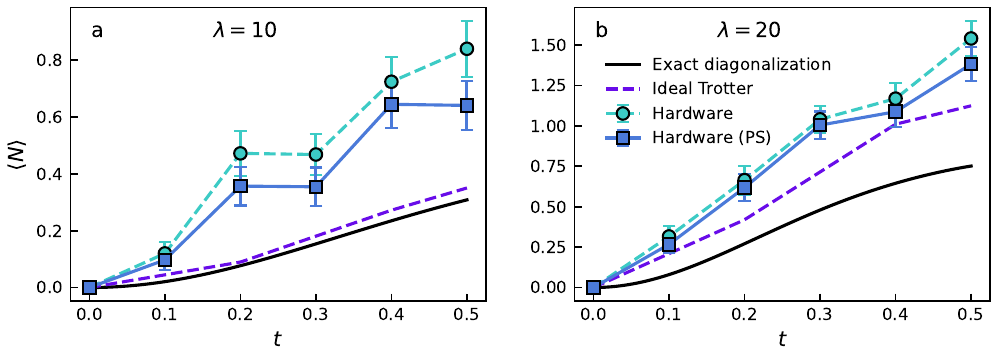}
\caption{
The total number operator expectation value, $\langle \sum_a N_a^{(\Lambda)} \rangle$ as a function of time for (a) $\lambda = 10$ and (b) $\lambda = 20$. The system is initialized in the free vacuum $\ket{0}^{\otimes n_Q}$ and then and evolved via exact diagonalization, ideal Trotterization, quantum hardware, and hardware with gauge-singlet post-selection. Error bars indicate one standard deviation, and the data correspond to truncation level $K=2$.
}
\label{fig:number_operator_vs_time}
\end{figure}

In addition to providing an indirect measure of simulation fidelity, the detected violations of the gauge singlet subspace can also be used to mitigate the errors by discarding non-singlet shots. We demonstrate this for the total occupation number operator, defined as $N_{\mathrm{tot}} = \sum_{a=1}^3 N_a$. The truncated number operator for a single oscillator, say the $x$ oscillator, is:
\begin{equation}
    N^{(\Lambda)}_x = a^{(\Lambda)\dagger}_x a^{(\Lambda)}_x = \sum_{i=0}^{K-1} 2^i \left( \frac{\mathds{1} - Z_i}{2} \right) \,.
\end{equation}
Here $Z_k$ denotes the Pauli $Z$ operator acting on qubit $k$, with identity on all other qubits. This operator is diagonal in the Fock basis, which coincides with the $Z$-measurement basis. Analogous expressions exist for the $y$ and $z$ oscillators. Each acts on a different set of qubits, so the total number operator is a linear combination of weight-1 $Z$ operators, plus an identity offset. Unlike the Loschmidt echo, this operator is not a projector onto a single basis state, and gauge singlet post-selection improves the noisy hardware result by discarding shots that violate the singlet condition. This is depicted in Fig.~\ref{fig:number_operator_vs_time}. Starting from the free vacuum at $t=0$, the expectation value grows from zero at a rate that increases with coupling strength. Post-selection reduces the error relative to exact diagonalization across all time steps by $16-38\%$ for $\lambda=10$ and $6-21\%$ for $\lambda=20$. 

\section{Discussion}
In this work, we considered the digital quantum simulation of arguably the simplest possible matrix model---a single Hermitian matrix with gauge group $\mathrm{SU}(2)$ and no fermions. Even for such an idealized system, several manipulations are required to translate the model into a form amenable to digital quantum simulation. First, the infinite-dimensional Hilbert space must be truncated to that of an $n_Q$-qubit system, and the Hamiltonian restricted to this subspace and expressed in terms of Pauli operators. Next, the time-evolution operator $e^{-i t H}$ is approximated using a Lie-Trotter formula and compiled into a quantum circuit. Finally, the circuit is executed on a noisy quantum device. Each step introduces error---Hilbert space truncation, Trotterization, and hardware noise. Although this general prescription is known~\cite{Gharibyan:2020bab, Rinaldi:2021jbg}, ours is the first study to implement it fully and systematically characterize the errors at each stage, culminating in execution on Quantinuum System Model H2 device.
Our analysis reveals that simulating the dynamics of matrix models represents a challenging task for current and near-term quantum hardware. Device noise levels limit the depth of the Trotter circuits that can be executed with tolerable fidelity. As circuit depth grows with the number of qubits, current hardware constraints restrict the truncation to Hilbert space dimensions that remain classically tractable---a limitation that, while precluding quantum advantage, is precisely what enables the systematic benchmarking carried out in this work. 

The depth limitation imposed by device noise also limits the duration and resolution of the time evolution. The Loschmidt echo, our primary observable of interest in this work, contains spectral information---in particular the energy gaps $\omega_{n_1 n_2} = E_{n_2} - E_{n_1}$---that may be extracted through signal processing. However, the frequency resolution $\Delta \omega = 2\pi / T$ and Nyquist frequency ${\omega_{\text{Nyq}} = \pi / \dd t}$ constrain which gaps are experimentally accessible. For our hardware parameters ($\dd t = 0.1$, $N_t = 5$), these correspond to $\Delta \omega \approx 10.5$ and $\omega_{\text{Nyq}} \approx 31.4$, limiting the range of resolvable frequencies even before accounting for noise. In practice, the short time series and hardware noise make meaningful spectral extraction infeasible. Longer time evolution---requiring either deeper circuits or improved error mitigation---would be needed to resolve spectral features on quantum hardware. A possible way forward that enables both longer time dynamics and larger truncations---resulting in significantly more terms in the Trotterized Hamiltonian evolution operator---is circuit compression~\cite{Zhang:2024kuf, DAnna:2025yzo, Gibbs:2025izl}, a classical optimization technique that builds a shorter depth ansatz circuit optimized to reproduce the unitary dynamics at late times.

In addition to characterizing these error sources, we investigated mitigation strategies. Whereas truncation and Trotter errors are intrinsic to the digital quantum simulation framework we utilized, device noise can be mitigated through various techniques. We were able to reduce error relative to the ideal Trotter simulation by up to 96\% through Zero-Noise Extrapolation (ZNE). We also introduced a gauge singlet post-selection scheme that exploits the fact that certain gauge-symmetry violations are signaled by odd occupation numbers in the $Z$-basis measurement outcomes, allowing them to be detected and discarded in post-processing.
The discard rate under this post-selection scheme characterizes hardware noise as a function of time and coupling; we find maximal discard rates of roughly 5-8\% for the final Trotter step, which corresponds to the deepest and thus noisiest circuits we ran. We also demonstrated that the post-selection improves expectation value accuracy -- for the case of the number operator, post-selection reduced the error by 6--38\% across time steps and coupling strengths. 

We note that both ZNE and gauge-singlet post-processing, as well as other error-mitigation techniques, can be expected to encounter difficulties as the system size (and thus the circuit dimensions) scale. In particular, the post-selection discard rate will likely grow to nearly 100\% in both the width and depth of the circuit. For this reason, we regard these and other mitigation strategies as a ``last-mile'' approach towards improving the accuracy of the simulation. Future work should focus on depth reduction through improved compilation and unitary synthesis, error suppression, and even partial error correction approaches.

It is important to note that our benchmarking is significantly aided by the fact that the quantum dynamics of a single Hermitian matrix are classically simulable. Firstly, for the qubit counts considered in this work, exact diagonalization remains tractable. Second, as discussed in Sec.~2 and App.~\ref{app:numerical_solution}, the model is exactly solvable, via fermionization for general $N$, or the radial reduction for $N=2$. This allows for classical simulability before truncating the system to a finite number of qubits. We regard this classical tractability as a \textit{feature} rather than a limitation for the present study. Our primary objective is to benchmark quantum simulation methodologies for matrix models on digital quantum hardware. The quartic interaction structure generates highly non-local couplings when expressed in the oscillator basis, resulting in deep quantum circuits. This makes matrix models qualitatively more challenging than commonly-studied spin systems. 

A primary motivation of this work is to build towards the digital quantum simulation of more complex and phenomenologically-interesting matrix models. A natural next target would be the mini-BFSS or mini-BMN models \cite{anous2019mini, han2020deep}, which involve $d=3$ Hermitian matrices, quartic commutator interactions, and fermions in the supersymmetric case. Even these ``mini'' models represent a big challenge for circuit compilation and hardware execution. More ambitiously, we look towards the true prize, supersymmetric $d=9$ BFSS and BMN matrix models. Ideally, digital quantum simulation could be used to validate and complement holographic approaches based on numerical solution of the supergravity equations relevant for the strongly coupled, large-$N$ regime (for instance, see the recent studies \cite{dias2025low, dias2025localized}). Advancing quantum simulation to these models requires surmounting important challenges. The supergravity description requires $N \to \infty$, yet the qubit resources for matrix model simulation scale as $\mathcal{O}(N^2)$, placing the holographic regime beyond near-term reach. The truncation level $K$ is also a consideration. The analysis of Sec.~\ref{sec:circuit} shows that, even for a single matrix with $N=2$, as truncation level $K$ increases, the number of Pauli terms in the decomposition grows exponentially. Fortunately, our analysis extends the finding of \cite{Rinaldi:2021jbg} that expectation values of gauge-invariant observables converge rapidly with increasing truncation. While this was demonstrated for static observables such as the Casimir operator and the energy, we show that it holds equally for the dynamical Loschmidt echo. More work is needed to understand the impact of the truncation on multi-matrix models.

\subsection*{Data and code availability}
The pre-computed numerical and experimental data, alongside the Python plotting scripts used to generate the figures in this work, are openly available on Zenodo \cite{hartnett_2026_20498411}.

\subsection*{Acknowledgments}
We thank Yuta Kikuchi and Matt DeCross for comments on the manuscript.
This research used resources of the Oak Ridge Leadership Computing Facility, which is a DOE Office of Science User Facility supported under Contract DE-AC05-00OR22725. Specifically, access to the Quantinuum device and emulator was provided by the HEP150 program, facilitated through the OLCF Quantum Computing User Program (QCUP).

\subsection*{Author Contributions} 
G.S.H. and E.R. conceived the original idea for the project and jointly developed the theoretical framework. G.S.H. wrote the software for the classical spectral methods and exact diagonalization. H.L. led the quantum circuit construction, compilation, and hardware execution, with E.R. providing the original Hamiltonian construction code and guidance and support for the hardware interface. Error mitigation strategies were divided between H.L., who implemented the zero-noise extrapolation (ZNE), and G.S.H., who developed the gauge-singlet post-selection scheme. All authors contributed to the drafting, reviewing, and editing of the final manuscript.

\bibliographystyle{JHEP}
\bibliography{refs}

\appendix

\section{
Numerical solution via spectral methods
\label{app:numerical_solution}}
In this Appendix, we detail our numerical approach for solving Eq.~\eqref{eq:radial_equation} using a Chebyshev spectral collocation method (for a general reference on spectral methods, see \cite{trefethen2000spectral}). The domain of the wavefunctions is the half-line $r \in [0, \infty)$, which is mapped to the unit interval via the coordinate transformation
\begin{equation}
    z = \tanh\left(\frac{r}{L}\right) \,, 
\end{equation}
where $L$ is a scale parameter chosen to match the characteristic length scale of the problem. This mapping sends $r = 0 \to z = 0$ and $r \to \infty \to z \to 1$.

On this computational domain, we employ Chebyshev polynomials as our basis functions, using the Chebyshev-Gauss-Lobatto collocation points
\begin{equation}
    z_j = \cos\left(\frac{\pi j}{N}\right) \,, \quad j = 0, 1, \ldots, N \,.
\end{equation}
The key advantage of spectral collocation is that derivatives can be computed exactly (in the space of polynomials of degree $\leq N$) using differentiation matrices $D_1$ and $D_2$. For a function represented by its values $f_j = f(z_j)$ at the collocation points (with the vector of such points denoted as $\mathbf{f}$), the first derivative is approximated as $(D_1 \mathbf{f})_j \approx f'(z_j)$. The differentiation matrix is given by
\begin{equation}
    (D_1)_{ij} = 
    \begin{cases}
        \displaystyle \frac{a_i}{a_j(z_i - z_j)} \,, & i \neq j \,, \\[0.8em]
        \displaystyle \sum_{k=0, \, k \neq i}^{N} \frac{1}{z_i - z_k} \,, & i = j \,,
    \end{cases}
\end{equation}
where $a_j = \prod_{k \neq j} (z_j - z_k)$. The second derivative matrix is simply $D_2 = D_1^2$.

To construct the Hamiltonian operator, we must express the second derivative $\dd^2/\dd r^2$ in terms of derivatives with respect to the computational coordinate $z$. Using the chain rule with $p(z) = \dd r/\dd z = L/(1-z^2)$, we obtain
\begin{equation}
    \frac{{\dd}^2}{\dd r^2} = \frac{1}{p} \frac{\dd}{\dd z} \left( \frac{1}{p} \frac{\dd}{\dd z} \right) \,.
\end{equation}
In matrix form, this becomes ${D_2^{(r)} = \text{diag}(1/\mathbf{p}) \, D_1 \, \text{diag}(1/\mathbf{p}) \, D_1}$, where the factors of $\text{diag}(1/\mathbf{p})$ account for the coordinate transformation. The discrete Hamiltonian matrix is then
\begin{equation}
    H = -\frac{1}{2} D_2^{(r)} + \text{diag}(\bm{V}) \,,
\end{equation}
where $\bm{V}_j = V(r_j)$, and $r_j$ are the physical coordinates corresponding to $z_j$.

Finally, we impose the boundary condition $u(0) = 0$ by removing the first row and column of $H$ (corresponding to $r = 0$), and we implicitly handle the asymptotic boundary condition $u(r \to \infty) = 0$ through the compactification—wavefunctions that decay sufficiently rapidly will naturally satisfy this condition. The eigenvalue problem $H \mathbf{u} = E \mathbf{u}$ is then solved using standard dense eigenvalue solvers to obtain the spectrum and eigenstates. 

\begin{figure}[H]
\centering
\includegraphics[width=\linewidth]{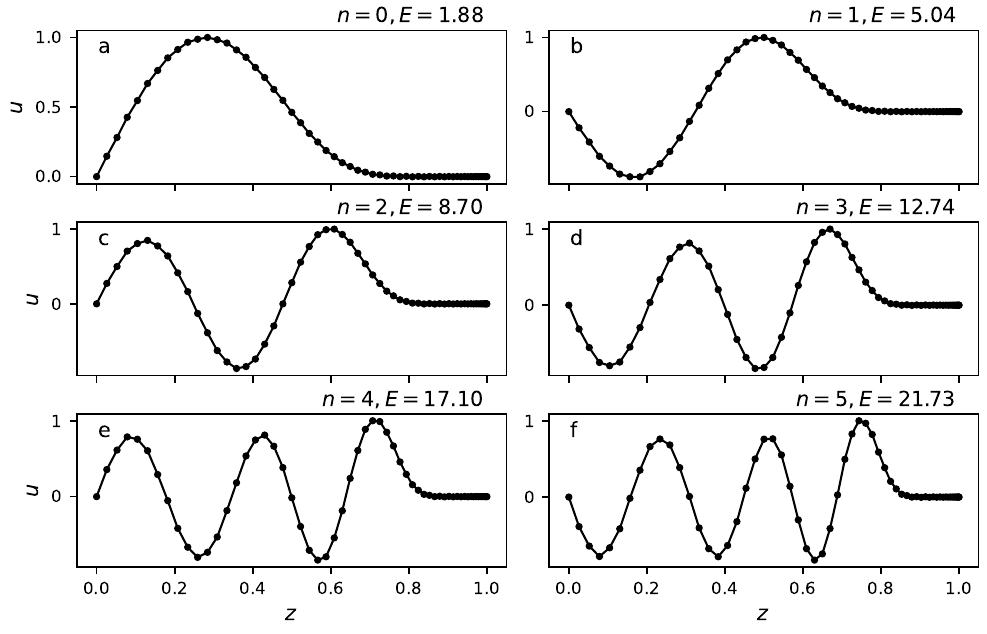}
\caption{The first 6 eigenfunctions of the time-independent radial Schr\"odinger equation, Eq.~\eqref{eq:radial_equation}, for parameters $\textrm{m}=1$ and $\lambda = 10$. The eigenfunctions are purely real, and are plotted as functions of the compactified coordinate $z = \tanh(r/L)$ for $L=3$ on a grid with $120$ discretization points.
}
\label{fig:radial_eigenfunctions}
\end{figure}

We now apply this method to compute the spectrum and eigenstates for several coupling strengths. Figure~\ref{fig:radial_eigenfunctions} shows the first six eigenfunctions for $\textrm{m}=1$ and $\lambda = 10$, and Tab.~\ref{tab:spectrum_schrodinger} presents the energy eigenvalues $E_n$ and overlap coefficients $|c_n|^2 = |\langle \psi_n | \mathrm{vac} \rangle|^2$ with the free vacuum for various coupling strengths. 

\begin{figure}[H]
\centering
\includegraphics[width=1\linewidth]{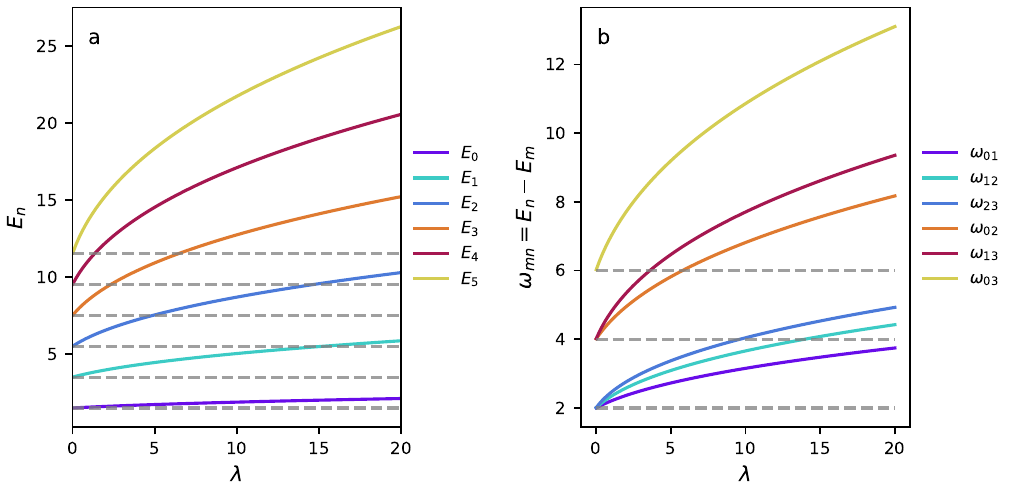}
\caption{
Spectrum of the interacting theory. 
(a) The first 6 energy eigenstates $E_n$, with $n=0, 1, ..., 5$. Dashed horizontal lines indicate the energies of the free theory,  Eq.~\eqref{eq:free_oscillator_spectrum}. 
(b) Spectral gaps for the first four eigenstates. At zero coupling, the spectrum reduces to that of a harmonic oscillator with $\omega_{mn} = 2(n-m)$, indicated by dashed horizontal lines.
}
\label{fig:energy_and_gap_vs_coupling}
\end{figure}

Figure~\ref{fig:energy_and_gap_vs_coupling} depicts the variation of the spectrum as a function of $\lambda$. Panel (a) shows the energies increasing monotonically with the coupling, with no level crossings---as expected, since the theory reduces to the spherically symmetric sector of a 3D anharmonic oscillator. Panel (b) shows the spectral gaps $\omega_{n_1 n_2}$ for the first four states. At zero coupling, the gaps reduce to $2\textrm{m} (n_2 - n_1)$ (we set $\textrm{m}=1$ in the plots). Since the gaps increase monotonically with $\lambda$, the lowest-frequency component in the Loschmidt echo is $\omega_{01} \geq 2 \textrm{m}$. This bounds the longest timescale that must be resolved to extract spectral information from device measurements.

\begin{table}[H]
\centering
\caption{
\label{tab:spectrum_schrodinger}
Low-lying energy spectrum and initial state overlaps for the quartic anharmonic oscillator. 
}
\setlength{\tabcolsep}{8pt}
\renewcommand{\arraystretch}{1.2}
\begin{tabular}{c cc @{\qquad} cc @{\qquad} cc}
\toprule
 & \multicolumn{2}{c}{$\lambda = 2$} & \multicolumn{2}{c}{$\lambda = 10$} & \multicolumn{2}{c}{$\lambda = 20$} \\
\cmidrule(lr){2-3} \cmidrule(lr){4-5} \cmidrule(lr){6-7}
$n$ & $E_n$ & $|c_n|^2$ & $E_n$ & $|c_n|^2$ & $E_n$ & $|c_n|^2$ \\
\midrule
0 & 1.603 & 9.94E-01 & 1.879 & 9.50E-01 & 2.114 & 9.01E-01 \\
1 & 3.966 & 5.14E-03 & 5.032 & 3.94E-02 & 5.859 & 7.09E-02 \\
2 & 6.544 & 5.39E-04 & 8.691 & 8.35E-03 & 10.28 & 1.98E-02 \\
3 & 9.300 & 3.31E-05 & 12.73 & 1.68E-03 & 15.21 & 5.59E-03 \\
4 & 12.21 & 2.30E-06 & 17.09 & 3.91E-04 & 20.55 & 1.77E-03 \\
\bottomrule
\end{tabular}
\end{table}

The numerical method can be verified against analytic and semiclassical limits. For $\lambda=0$, the system reduces to the odd sector of the simple harmonic oscillator, corresponding to $E_n = (2n + 3/2) \, \textrm{m}$. Our spectral solver reproduces these energies with relative errors of $0.02-0.1\%$. For $\textrm{m}=0$, comparison with the WKB approximation~\cite{bender2013advanced} yields relative errors decreasing from $\sim 1\%$ to $0.06\%$ as $n$ increases from $0$ to $7$, consistent with asymptotic WKB validity. 

While spectral methods typically achieve exponential convergence for analytic functions, the exponential decay of wavefunctions at large $r$ renders $u$ non-analytic at the compactified boundary $z \to 1$, limiting convergence to algebraic rates. The asymptotic behavior for $\lambda = 0$ is
\begin{equation}
    \label{eq:asymptotic_free}
    u(r) = A \, \frac{e^{-\textrm{m} r^2/2}}{\sqrt{r}} \left( 1 + \mathcal{O}(r^{-2}) \right) + B \, \frac{e^{\textrm{m} r^2/2}}{\sqrt{r}} \left( 1 + \mathcal{O}(r^{-2}) \right)\,,
\end{equation}
while for $\lambda \neq 0$,
\begin{equation}
    \label{eq:asymptotic_interacting}
    u(r) = A \, \frac{e^{-\frac{2 \sqrt{2} \textrm{m}^2 r}{\sqrt{\lambda }}-\frac{\sqrt{\lambda } r^3}{12
   \sqrt{2}}}}{r} \left(1 + \mathcal{O}(r^{-1}) \right) + B \, \frac{e^{\frac{2 \sqrt{2} \textrm{m}^2 r}{\sqrt{\lambda }} + \frac{\sqrt{\lambda } r^3}{12
   \sqrt{2}}}}{r} \left(1 + \mathcal{O}(r^{-1}) \right) \,,
\end{equation}
where $A, B$ are integration constants fixed by boundary conditions. We require $B=0$ for square-integrability. Factoring out the asymptotic form could restore exponential convergence, but the current accuracy is sufficient for our purposes.

\section{Fock-space truncation and qubit encoding: further details \label{app:truncation}}
The Hilbert space of a traceless $N \times N$ Hermitian matrix is infinite-dimensional, posing an immediate challenge for digital quantum simulation. To represent the quantum state using a finite number of qubits, the Hilbert space must be truncated. We use the truncation scheme introduced in~\cite{Gharibyan:2020bab}, where the states of the $n_Q$-qubit system correspond to the first $2^{n_Q}$ states of the non-interacting infinite-dimensional system. This may be accomplished in two steps. First, the matrix $X$ is expanded into $N^2 - 1$ oscillator degrees of freedom, one for each generator of $\mathrm{SU}(N)$. This was done in Sec.~\ref{sec:truncation}. Second, each oscillator’s Hilbert space is truncated to finite dimension.

The truncation scheme is formulated in the Fock basis of the free theory. The oscillator representation makes explicit that the full Hilbert space (before gauging) is a Fock space spanned by states $\ket{\bm{n}} = \ket{n_1, \ldots, n_{N^2-1}}$, where $n_a \in \mathbb{N}$ denotes the occupation number of the $a$-th oscillator, associated with each $\mathrm{SU}(N)$ generator. The usual ladder operators are used to transition from one Fock state to another:
\begin{equation}
    a_a = \sqrt{\frac{\textrm{m}}{2}} X_a + \frac{i}{\sqrt{2 \textrm{m}}} P_a \,, \quad
    a_a^{\dagger} = \sqrt{\frac{\textrm{m}}{2}} X_a - \frac{i}{\sqrt{2 \textrm{m}}} P_a \,.
\end{equation}
The free (quadratic) part of the Hamiltonian, $H_0 = \textrm{m} \sum_a \left( N_a + \frac{1}{2} \right)$, is diagonal in this basis, with $N_a = a_a^\dagger \, a_a$ the per-mode number operator.

For each oscillator $a = 1, \ldots, N^2-1$, we truncate the infinite-dimensional Fock space and retain only the first $\Lambda$ states ($\ket{0}_a, \ket{1}_a, \ldots, \ket{\Lambda - 1}_a$), corresponding to the lowest energy eigenstates of the free Hamiltonian $H_0$. The truncated Hilbert space is then
\begin{equation}
    \mathcal{H}_{\text{trunc}} = \bigotimes_{a=1}^{N^2-1} \text{span}\{\ket{0}_a, \ket{1}_a, \ldots, \ket{\Lambda-1}_a\} \,,
\end{equation}
which has dimension $\Lambda^{N^2-1}$. For implementation on qubit-based quantum computers, we set $\Lambda = 2^K$, allowing each oscillator to be encoded using $K$ qubits. The total number of qubits required is then
\begin{equation}
    n_Q = (N^2 - 1) K \,.
\end{equation}

Next, we explain how to truncate operators. The Fock states are encoded via the binary representation
\begin{equation}
    \ket{j}_a = \ket{b_{K-1}}_a \ket{b_{K-2}}_a \cdots \ket{b_0}_a \,, \quad \text{with} \quad j = \sum_{\ell=0}^{K-1} b_\ell \, 2^\ell \,,
\end{equation}
and $b_\ell \in \{0,1\}$. We then introduce truncated ladder operators:
\begin{equation}
    \label{eq:truncated_ladder_ops}
    a_a^{(\Lambda)} \ket{k}_a =
    \begin{cases}
        \sqrt{k} \ket{k - 1}_a & \text{if } 1 \leq k < \Lambda \\
        0 & \text{otherwise}
    \end{cases} \,, \quad
    (a_a^{(\Lambda)})^{\dagger} \ket{k}_a =
    \begin{cases}
        \sqrt{k + 1} \ket{k + 1}_a & \text{if } 0 \leq k < \Lambda - 1 \\
        0 & \text{otherwise}
    \end{cases} \,.
\end{equation}
These act like the usual ladder operators within the truncated space. They may be expressed in terms of Fock state transitions such as $\ket{j+1}_a \bra{j}_a$, which in turn can be written as a tensor product of single-qubit operators:
\begin{equation}
    \ket{j+1}_a \bra{j}_a = \bigotimes_{\ell=0}^{K-1} \ket{b'_\ell}_a \bra{b_\ell}_a \,,
\end{equation}
where $b'_\ell$ denotes the $\ell$-th bit of $j+1$.  Finally, the single-qubit projectors and transition operators are expressed in terms of Pauli matrices as
\begin{equation}
    \ket{0}\bra{0} = \frac{\mathds{1} - Z}{2} \,, \quad
    \ket{1}\bra{1} = \frac{\mathds{1} + Z}{2} \,, \quad
    \ket{0}\bra{1} = \frac{X + i \, Y}{2} \,, \quad
    \ket{1}\bra{0} = \frac{X - i \, Y}{2} \,.
\end{equation}

Having established truncated versions of the ladder operators, the procedure for truncating any operator $O$ is then: express $O$ in terms of ladder operators, then replace those with the truncated ladder operators. The resulting truncated operator will be denoted $O^{(\Lambda)}$. For example, the truncated per-mode number operator is $n_a^{(\Lambda)} = (a_a^{(\Lambda)})^{\dagger} a_a^{(\Lambda)}$. This scheme enables all operators to be expressed as linear combinations of $n_Q$-qubit Pauli strings as in Eq.~\eqref{eq:operator_pauli_decomposition}, where $\mathcal{P}_n$ is the $n_Q$-qubit Pauli group, $P$ are individual Pauli elements, and $c_P$ is the (in general complex) coefficient of the Pauli term $P$. This decomposition allows both the Trotterized time evolution circuits to be constructed and the observables of interest to be measured.

An important consequence of the truncation is that the truncated ladder operators do not obey the standard commutation relations, so that 
\begin{equation}
    [a_a^{(\Lambda)}, a_b^{(\Lambda)\dagger}] = \delta_{ab} \left( \mathds{1} - \Lambda \ket{\Lambda-1}_a \bra{\Lambda-1}_a \right) \,, \quad \text{(no sum on $a$)}
\end{equation}
This introduces an ambiguity in the operator truncation prescription. The natural prescription, which we utilize here, is to normal-order all operators before truncation, placing creation operators to the left of annihilation operators.

The modified commutation relation also has consequences for gauge symmetry. The truncated gauge generators $G_a^{(\Lambda)}$ remain well-defined operators, but they no longer exactly satisfy the $\mathrm{SU}(N)$ algebra. Similarly, they do not generate exact gauge transformations on the truncated position operators, and they fail to commute with the truncated Hamiltonian: $[G_a^{(\Lambda)}, H^{(\Lambda)}] \neq 0$. As a result, eigenstates of $H^{(\Lambda)}$ are not exact gauge singlets. However, these violations are localized to states with significant support on the highest occupation numbers and are exponentially suppressed for low-lying states at moderate values of $\Lambda$~\cite{Rinaldi:2021jbg}.

\end{document}